\documentclass[aps,pra,10pt,amsmath,amsfonts,amssymb,twocolumn,superscriptaddress]{revtex4}
\usepackage{amssymb,amsfonts}
\usepackage{graphicx,psfrag,color}
\usepackage{dcolumn}
\usepackage{bbm}
\usepackage{mathbbol}
\usepackage{scalerel}
\usepackage{braket}
\usepackage{amsmath}
\usepackage[TS1,T1]{fontenc}
\usepackage{textcomp}
\usepackage[toc,page]{appendix} 
\usepackage[utf8]{inputenc}

\newcommand{\up}{\uparrow}

\newcommand{\down}{\downarrow}
\newcommand*{\balancecolsandclearpage}{%
  \close@column@grid
  \clearpage
  \twocolumngrid
}

\begin{document}

\title{Quantum Corralling}

\author{Rafael Vieira}
\author{Gustavo Rigolin}
\email{rigolin@ufscar.br}
\affiliation{Departamento de F\'isica, Universidade Federal de
S\~ao Carlos, 13565-905, S\~ao Carlos, SP, Brazil}
\author{Edgard P. M. Amorim}
\email{edgard.amorim@udesc.br}
\affiliation{Departamento de F\'isica, Universidade do Estado de
Santa Catarina, 89219-710, Joinville, SC, Brazil}

\date{\today}

\begin{abstract}
We propose a robust and efficient way to store and transport quantum information via one-dimensional 
discrete time quantum walks. We show how to attain an effective dispersionless wave packet evolution
using only two types of local unitary operators (quantum coins or gates), 
properly engineered to act at predetermined times and at specific lattice sites during the system's time evolution. In particular, we show that 
a qubit initially localized about a Gaussian distribution 
can be almost perfectly confined during 
long times 
or sent hundreds lattice sites away from its original location 
and later almost perfectly reconstructed
using only Hadamard and $\sigma_x$ gates.
\end{abstract}


\maketitle

\section{Introduction} 

In quantum mechanics the ``spreading'' of the wave function 
is a ubiquitous characteristic of time-independent Hamiltonians \cite{bal98,gre00}. 
In this scenario, our knowledge of where a microscopic particle is located 
diminishes as time goes by. This natural delocalization of quantum entities is
most of the time a hindrance to an efficient 
implementation of quantum communication protocols
or to the proper execution of a quantum computational task.  Indeed,
the main problem of any type of communication is the ability to recover 
with a reasonable level of fidelity the information content sent from one point to 
another \cite{shannon1948mathematical} and it is thus important to know where
a quantum particle is located if we want to employ it as a carrier of information.
In other words, we have to somehow reduce or suppress the unwanted spreading of the wave function of a quantum particle to use it efficiently 
as a carrier of information \cite{hua21}.

Here we pose and solve the above problem in the framework of one-dimensional discrete time quantum walks \cite{aharonov1993quantum, kempe2003quantum}. 
We show a very simple and robust scheme allowing 
an effective dispersionless time evolution of a wave package describing the position probability density of 
a qubit (our quantum  walker). This scheme
can also be adapted to efficiently store the 
information content of a quantum state at a given location. Actually, as we will see,
we can dispersionlessly send the information, store it at another place, 
and send it again, repeating
the previous steps many times, 
without disturbing the information carried by the qubit.

The present protocol is inspired and works similarly in spirit to the techniques 
employed by cowboys and cowgirls to herd or drive livestock from one place to another 
inside a cattle handling facility. This process, usually called ``corralling'', aims to either send cattle from one place to another or corral it in a cattle pen. 
In order to drive the cattle from one place to another,
a cowgirl closes a given gate while opening another one at the right place and time.
By successively closing and opening gates at the right times and places, she can
smoothly move the cattle from one location to another. 
The timing of the opening and closing of the gates has to be precise, 
otherwise the cattle will disperse or even move in the wrong direction. 
As we will see here, the protocol we describe and call ``quantum corralling'' works \textit{mutatis mutandis} exactly in the same way. 
The opening and closing of gates in a cattle handling facility are 
substituted by the activation or inactivation of $\sigma_x$ gates at right times and places while the untamed cattle is a Gaussian wave packet
being herded by the unitary evolution of quantum mechanics, which is not a very good herder
due to its inherent dispersive aspect. See Figs. \ref{fig1} to
\ref{fig3} and specially the videos available online as 
Supplementary Material to this work for a visual and easy understanding of the quantum corralling protocol.

\section{The protocol's platform} 

Quantum walks are a promising framework to implement a 
variety of quantum tasks, such as quantum search algorithms \cite{shenvi2003quantum,tulsi2008faster} and universal quantum computation \cite{childs2009universal,lovett2010universal}. A rich  dynamical behavior 
can be engineered in a quantum walk, ranging from diffusive to ballistic transport 
\cite{vieira2013dynamically,vieira2014entangling,orthey2019connecting,li2013position,
orthey2017asymptotic,ghizoni2019trojan,cardano2015quantum,su2019experimental}, 
and many physical systems can be used to
experimentally build a quantum walk \cite{wang2013physical,venegas2012quantum}. 
See also Refs. \cite{bos03,chr04,alb04,nik04,sub04,osb04,ple04,sem05,chr05,shi05,woj05,li05,chi05,kar05,
har06,huo08,gua08,ban10,kur11,god12,apo12,lor13,sou14,hor14,shi15,lor15,zha16,che16,nic16,est17,est17a,alm17,alm18,alm19} for quantum 
state transfer protocols using other platforms such as  
spin chains or continuous variable systems.

For our purposes, we can think of a quantum walker as a spin-$1/2$ particle (qubit) 
placed on a regular one-dimensional lattice where each site represents a discrete position. Its dynamical evolution is driven by a unitary operator formed by a quantum coin (gate) 
and a conditional displacement operator. The quantum coin acts on the qubit
changing
its spin state and the displacement operator moves the up (down) spin state to the right (left) adjacent position. The interference between the up spin state moving to 
the right with the down spin state moving to the left is 
the reason for the rich dynamics of quantum walks. Depending on the initial state and on 
the coin, we either get localization or transport of the wave function, with 
the latter being diffusive or ballistic.

\subsection{Mathematical formalism}

We now highlight the main features of the formalism and notation fully developed in 
Refs. \cite{vieira2013dynamically,vieira2014entangling} that are needed for our 
present purposes. For further details we direct the reader to those aforementioned references.

The internal degree of freedom of the quantum walker, 
for instance the spin of an electron or the polarization of a photon, and its external 
degree of freedom (position) are described by the Hilbert space $\mathcal{H}=\mathcal{H}_C\otimes\mathcal{H}_P$. Here $\mathcal{H}_C$ is the coin space, a complex two-dimensional vector space spanned by the vectors $\left\{\ket{\up}, \ket{\down}\right\}$, and 
$\mathcal{H}_P$ is the position space, a numerable infinite dimensional vector space 
spanned by $\left\{\ket{j}\right\}$, with $j$ being an integer denoting the 
discrete position of the walker on a one-dimensional lattice. We assume that 
the information content
of the walker is encoded in its internal degree of freedom. 

An arbitrary initial state where the internal degree of freedom is not entangled with 
the position degree of freedom can be written as
\begin{equation}
\ket{\Psi(0)}=\left[\cos\alpha\ket{\up}+e^{i\beta}\sin\alpha\ket{\down}\right]\otimes\sum_jf(j)\ket{j}, \label{state0}
\end{equation}
where we sum over all integers $j$, $\alpha\in[0,\pi/2]$ and $\beta\in[0,2\pi]$. Note that for simplicity we set $\alpha$ to be half the polar angle $\theta$ 
in the Bloch sphere representation ($\theta \in[0,\pi]$) while $\beta$ is the usual azimuthal angle. Since we will be dealing with a Gaussian wave packet initially centered
at the origin,
\begin{equation}
f(j)=Ae^{-[j^2/(4 s^2)]}/(2\pi s^2)^{1/4}.
\end{equation}
Here $s$ is the initial standard deviation and $A$ is a normalization constant to
guarantee that $\sum_j|f(j)|^2=1$. If $j$ were a continuous variable we would have $A=1$.
Eventually, in our numerical experiments, we will set $|j|\leq j_{max}$ and 
$A$ will be chosen to guarantee the normalization condition in this scenario. 

Our main goal is to tune the time evolution of the system such that at a 
chosen position $j=x$, we will have at time $t$ the same wave packet we had 
at $t=0$ but now centered at $x$. We want $|\Psi(t)\rangle$ to be 
$|\Psi(0)\rangle$ displaced to position $x$. In other words,
we want $|\Psi(t)\rangle = D_x|\Psi(0)\rangle$, 
where $D_x = \mathbb{1}_C\otimes \sum_j |j+x\rangle\langle j|$ and 
$\mathbb{1}_C$ is the identity operator in the coin space. In this case 
we will achieve an effective dispersionless time evolution 
that preserves the information encoded
in the spin state: the $t=0$ spin state 
$\cos\alpha\ket{\up}+e^{i\beta}\sin\alpha\ket{\down}$ will be the spin state
at $j=x$ and time $t$.

The walker's state after $n$ discrete time steps is given by \cite{vieira2013dynamically,vieira2014entangling}
\begin{equation}
\ket{\Psi(n)}=\mathcal{T}\prod_{t=1}^{n} U(t)\ket{\Psi(0)}, 
\label{evolution}
\end{equation}
where $\mathcal{T}$ indicates a time-ordered product and 
\begin{equation}
U(t)=S\mathcal{C}(t), 
\end{equation}
where
\begin{equation}
S=\sum_j(\ket{\up}\bra{\up}\otimes\ket{j+1}\bra{j} 
+\ket{\down}\bra{\down}\otimes\ket{j-1}\bra{j}) 
\end{equation}
is the conditional displacement operator, 
moving a spin up (down) to the right (left), and
\begin{equation}
\mathcal{C}(t) = \sum_jC(j,t)\otimes |j\rangle\langle j|, 
\end{equation}
where $C(j,t)$ is the coin
operator that acts on the internal degree of freedom at position $j$ and at the time $t$.
Note that in general $C(j,t)$ depends on both $j$ and $t$ and only if we have the 
same coin at all sites we get  
$\mathcal{C}(t) = C(t)\otimes \mathbb{1}_P$, where $\mathbb{1}_P$ is the identity operator
in the position space.
In this work we only use two coins, the Hadamard gate, $H=(\ket{\up}\bra{\up}+\ket{\up}\bra{\down}+\ket{\down}\bra{\up}-\ket{\down}\bra{\down})/\sqrt{2}$, and the 
not gate, $\sigma_x=\ket{\up}\bra{\down}+\ket{\down}\bra{\up}$.

\subsection{Pure Hadamard dynamics} 

If in all sites we only have the Hadamard coin,
namely, $\mathcal{C}(t) = H\otimes \mathbb{1}_P$, a Gaussian state centered at the origin
as given by Eq.~(\ref{state0}) and evolving according to Eq.~(\ref{evolution}) will 
split into two dispersive Gaussian wave packets, 
one moving to the left and the other moving to the right. Numerical analysis proves the
latter claim for small and large values of the initial standard deviation $s$. 
For large enough $s$ and small times, we show in the Appendix \ref{ap0} that we essentially have
\begin{eqnarray} 
\ket{\Psi(t)}&=&\ket{\psi_R}\otimes \sum_jf(j-t/\sqrt{2})|j\rangle \nonumber \\
&& + (-1)^t\ket{\psi_L}\otimes\sum_jf(j+t/\sqrt{2})|j\rangle,
\label{Psit}
\end{eqnarray}
where $\ket{\psi_L}$ and $\ket{\psi_R}$ are orthogonal 
states that depend
only on $\alpha$ and $\beta$. It is clear from Eq.~(\ref{Psit})
that the splitting of the initial Gaussian into two oppositely moving ones 
occurs regardless of the initial condition of the internal degree of 
freedom. Of course, for larger times the dispersion of the wave packets can no longer be 
ignored and the approximation above no longer applies.

We also realize looking at Eq.~(\ref{Psit}) that 
when only Hadamard coins are present, we cannot get a dispersionless wave packet evolution
(the original wave package split into two wave packages).
In order to ``tame'' and properly drive the wave packet in the direction we want and 
to obtain an effective dispersionless evolution, we need to add ``gates'' at specific
places and leave them ``closed'' during a certain time interval. As we show next, this
is achieved by exchanging Hadamard coins to $\sigma_x$ coins at certain lattice points
(closing the gate) and then later, at an appropriate time, changing back to  
Hadamard coins (opening the gate). By proceeding in this way, we will be able to  
corral the wave packet.

\subsection{Fidelity} 

Before we present more technically the protocol, it is important to
define the figure of merit we will be using to verify whether or not the information 
content encoded in the internal degree of freedom was stored or transmitted flawlessly.
We quantify the similarity between the evolved state at time $t$ with the initial one
at $t=0$ by computing the fidelity between those two states: $F(t)=|\bra{\Psi(0)}D_x^\dagger\ket{\Psi(t)}|^2$, where $D_x|\Psi(0)\rangle$ is the initial state displaced to $x$,
the center of the wave packet given by $|\Psi(t)\rangle$, and $D_x^\dagger$ is
the adjoint of $D_x$. If $F=1$ the two states
are the same up to an overall phase and if $F=0$ they are orthogonal.

\section{The protocol} 

Let us start showing how to keep a Gaussian state confined or
corralled. Later, we will explain how to drive this Gaussian state from one place to another.
As outlined above, corralling is achieved by changing the Hadamard gate to the
$\sigma_x$ gate at specific lattice sites. The idea behind using the $\sigma_x$ gate 
is related to the fact that it acts as a not gate, changing up (down) spin states to
down (up) spin states. As such, when we apply the conditional displacement after the 
action of this coin we will reverse the movement of the qubit. The $\sigma_x$ gate 
effectively ``blocks'' the passage of the qubit (similarly to 
the act of closing the gate while corralling cattle).

Being more specific, if initially the center of the Gaussian wave packet is located at
$j_{c} =(l+r)/2$ 
and the left and right blocking gates are placed at positions $l$ and $r$, 
respectively, we must set 
\begin{eqnarray}
C(j,t)&=&
\Bigg\{\hspace{-.5cm}
\begin{array}{c}
\sigma_x,\text{ if } j=l,  \\
\hspace{0.6cm}H,\text{ if } l<j<r, \\
\sigma_x,\text{ if } j=r.
\end{array}
\label{Hcoin}
\end{eqnarray}
Numerical analysis shows that we can get an almost flawless corralling for considerably 
long times (of the order of thousands of time steps) and without affecting the Gaussianity 
of the wave packet if 
the $\sigma_x$ gates are placed at or further than 
three standard deviations from the wave packet's center.
This means that $l\leq j_c - 3s$ and $r\geq j_c + 3s$. 
Moreover, the analytical result reported in the Appendix \ref{ap0} also shows that 
the greater the initial Gaussian width the more efficient is the present protocol. 
In other words, the greater the initial standard deviation the greater the 
fidelity of the corralled state at a given fixed time. This can be understood at the 
light of the Heisenberg uncertainty principle. A very narrow Gaussian in the position
space implies a greater dispersion in momentum, which inevitably leads to a faster spreading of the wave packet. It is worth noting that
numerical analysis shows that the efficiency of the present protocol 
decreases as the Gaussians become narrower. 
As we decrease the dispersion in position of the initial
wave package, we will reach a threshold below which the present protocol cannot achieve
a nearly perfect transmission (see Appendix \ref{apB} for details).  

\begin{figure}[!htb]\begin{center}
\includegraphics[width=8cm]{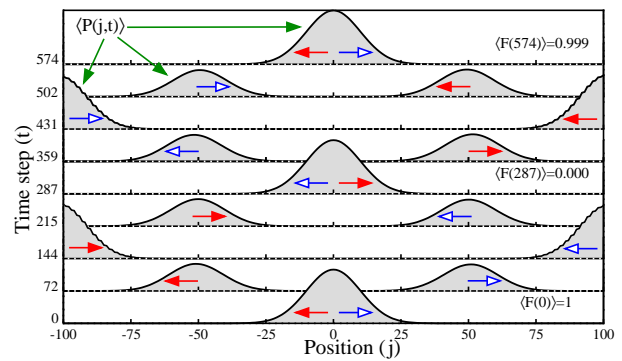}
\caption{(color online). Average probability distribu\-tion $\langle P(j,t) \rangle$ 
as a function of position $j$ and 
time step $t$. Here $P(j,t) = |(\bra{\uparrow}\bra{j})\ket{\Psi(t)}|^2
+|(\bra{\downarrow}\bra{j})\ket{\Psi(t)}|^2$. We used 
$451$ different spin initial conditions to compute the averages.  
The red and blue arrows indicate the direction of movement of the two Gaussian wave packets that split from the original
one due to the dynamics of the system (Hadamard walk). 
The average fidelities at specific times are also
shown. See text for details. Here and in all graphics all quantities are 
dimensionless.} 
\label{fig1}
\end{center}\end{figure}

As a concrete illustration of what we just said, we show in Fig.~\ref{fig1} the average
results of several numerical experiments using a Gaussian state with a fixed standard deviation $(s=10)$ and hundreds of different spin initial conditions. 
We work with $451$ different initial qubit states and following the notation given in Eq.~(\ref{state0}) we pick a representative sample of values for $\alpha$ and 
$\beta$ that covers their entire range. 
We start at their lowest values and generate in increments of $\pi/20$ the remaining ones, all the way up until we reach their upper bounds. Then we work with all combinations of the 
previously generated values of $\alpha$
and $\beta$ as our initial conditions. 
This is how we get the $451$ cases of different
initial qubit states. We center the Gaussian at $j_c=0$ and insert the $\sigma_x$ gates at
$l=-101 $ and $r=101$. 

In Fig.~\ref{fig1} the time starts at $t=0$ and after $t=287$ steps 
the wave package returns to 
its initial position, giving the same probability distribution in position 
space. 
However, due to the  reflection of the two split wave packages at the $\sigma_x$ gates
and to the dynamics associated with the Hadamard walk, 
the spin state when the divided wave packets first
meet is orthogonal to the initial one. We need to wait another round 
to get the same global state, where another relative phase shift of $\pi$ between
the up and down states compensates the first one. 
Therefore, at $t=574$ steps the system returns almost exactly to the original initial state. At this time we get an average fidelity of $0.999$. 

It is worth mentioning that the previous average fidelity is computed
by averaging the fidelities associated with all the $451$ different initial conditions
described above.
This means that the corralling works very well independently of the initial condition ascribed to the internal degree of freedom. Although not shown here, we checked the distribution for the 
fidelities of all the $451$ numerical experiments and we observed that all of them lie very close to the average value, corroborating the independence of the reported results on the initial spin state. 
Also, it is important to measure 
the quantum state at the right time. If we measure the state one step before or after
the right time, we get zero fidelity. This is due to the $(-1)^t$ term appearing in Eq.~(\ref{Psit}). A measurement in an odd time leads to $(-1)^t=-1$, which is 
equivalent to a phase shift of $\pi$ and thus to the measured state being orthogonal 
to the initial one.

\subsection{Single shot herding} 

Let us now move to the description of how to corral 
a Gaussian wave packet from one place to another,
attaining at the end an effective dispersionless transmission.
In this scenario there are two classes of
protocols. The first one drives the Gaussian state from one corral to another in a single shot, without interrupting 
the driving process along the way. The second class
of protocols is such that before reaching its final destination, the Gaussian wave packet is provisionally corralled in
one or several intermediate corrals.

\begin{figure}[!htb]
\includegraphics[width=8cm]{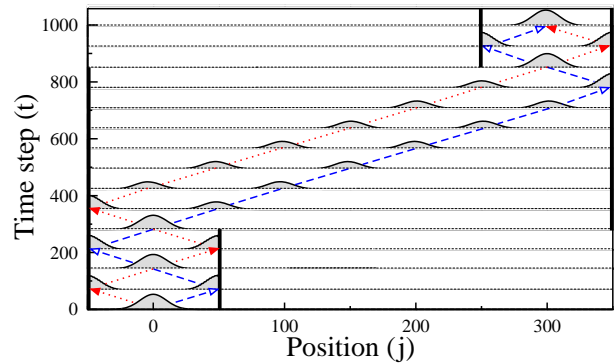}
\caption{(color online). Average probability distribution $\braket{P(j,t)}$ 
as a function of the position $j$ and time $t$
using the same $451$ initial conditions 
employed in Fig.~\ref{fig1}. Contrary to Fig.~\ref{fig1}, we now employ corrals whose 
gates are five standard deviations away from the initial wave package's center.
The right gate ($j=50$) is opened 
($\sigma_x$ coin changed to a Hadamard coin) when the center of both split wave packages 
coincide after two reflections at the gates of the corral ($t=282$). 
At this time we can already close the far right
gate of the new corral to which we will be corralling the Gaussian state.
Subsequently, half of the wave package moves to the right (dashed-blue line) and the other 
half, after reflecting again at the left gate of the old corral, also starts moving to the right (dotted-red line).  The confinement at the new corral is achieved 
by closing its left gate ($j=250$)
after the two wave packages' center meet for the first time
inside it ($t=849$). The average fidelity for this process at $t=995$ is 
$\braket{F(t)}=0.998$, which implies an almost flawless transmission. 
}
\label{fig2}
\end{figure}

Both protocols are built on slight modifications of the previous
corralling protocol, whose goal was to keep a Gaussian state confined indefinitely at a given corral. For the single
shot protocol, we can drive the Gaussian state to the right if at the appropriate time 
we exchange the $\sigma_x$ coin with the Hadamard coin at
the far right of the corral (we open the gate of the corral). In this way the original Gaussian
wave package will move to the right in two separated wave packages, which will be corralled at another location.
This is achieved by exchanging at the right time a Hadamard coin with a $\sigma_x$ coin 
at the far left of the new corral 
where we want to keep the Gaussian wave package confined (we are closing the gate of the corral now). See Fig.~\ref{fig2}
for details.

\subsection{Multiple station herding} 

\begin{figure}[!htb]
\includegraphics[width=8cm]{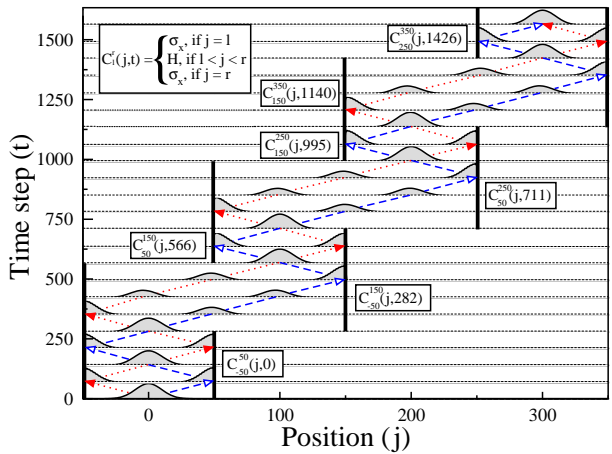}
\caption{(color online). Same as Fig.~\ref{fig2} but now we have several intermediate corrals of the 
same size. The superscript $r$ and the subscript $l$ defining the coin operator  
$C^r_l(j,t)$ keep track of the location of the $\sigma_x$ coins. In the graphics above,
the several values for the time within a given coin operator determine when 
it is activated and the previous one deactivated. For instance, 
$C_{-50}^{50}(j,0)$, $C_{-50}^{150}(j,282)$, and $C_{50}^{150}(j,566)$ imply 
that from $t=0$ to $t<282$ we have $\sigma_x$ coins (closed gates) at $j=-50$ and $j=50$ with the remaining sites given by
Hadamard coins. From $t=282$ to $t<566$, 
the only sites where we have $\sigma_x$ coins are
at $j=-50$ and $j=150$. At $t=566$, the $\sigma_x$ coins are only acting on sites 
$j=50$ and $j=150$. In a similar way we should read the remaining coin operators
shown in the graphics.
At the time $t=1566$, we have an average fidelity  given by $0.997$. Again, an almost
flawless transmission.}
\label{fig3}
\end{figure}

This class of protocols is built by successive applications of the previous one, where 
the right gate of the previous corral is the left gate of the next one. 
After reaching a given
corral, we repeat the single shot protocol, opening and closing the appropriate gates of the old and new corrals as explained above. 
Note that here we can also keep a Gaussian wave package during different times at different corrals. Furthermore, the 
multiple station protocol can be used to drive the qubit to a given place where a quantum gate can act upon it. In this way,
going back and forth to multiple corrals, where different quantum gates are installed, we can implement a variety of quantum 
computational tasks. See Fig.~\ref{fig3} for all the details. 

\section{Disorder} 

In order to investigate the robustness of the quantum corralling 
protocol in a more realistic scenario, 
we will analyze its response to slight variations about the optimal settings leading
to the almost perfect transmissions reported above. 
We will introduce errors (disorder) in
the quantum coins needed to implement the quantum walk's dynamics. And for definiteness,
from now on we will work with a fixed initial spin state, namely, 
$(\ket{\up}+i\ket{\down})/\sqrt{2}$, and we will focus on the 
multiple station protocol, whose operation is more prone to be affected by 
disordered quantum coins.

An arbitrary coin can be written as \cite{vieira2013dynamically,vieira2014entangling}
\begin{eqnarray}
C(j,t)&=&
\sqrt{q(j,t)}\ket{\up}\bra{\up} + \sqrt{1-q(j,t)} e^{i \theta(j,t)}\ket{\up}\bra{\down} 
\nonumber \\
&&+\sqrt{1-q(j,t)} e^{i \phi(j,t)} \ket{\down}\bra{\up}\nonumber \\
&&-\sqrt{q(j,t)}e^{i [\theta(j,t)+\phi(j,t)]}\ket{\down}\bra{\down},
\end{eqnarray}
where $0 \leq q(j,t) \leq 1$ and $-\pi\leq \theta(j,t), \phi(j,t) \leq\pi$. 
In this notation, the Hadamard coin is such that $q=1/2$ and $\theta=\phi=0$ 
while for the $\sigma_x$ coin we have $q=\theta=\phi=0$. 

We introduce 
disorder in a given coin by the following prescription  \cite{vieira2013dynamically,vieira2014entangling,vie18,vie19,vie20},
\begin{eqnarray}
q(j,t_n)=|q(j,t_{n-1})+\delta q(j,t_n)|,\nonumber\\
\theta(j,t_n)=\theta(j,t_{n-1})+\pi\delta \theta(j,t_n),\nonumber\\
\phi(j,t_n)=\phi(j,t_{n-1})+\pi\delta \phi(j,t_n), \nonumber
\end{eqnarray}
where $\delta q(j,t_n), \delta \theta(j,t_n)$, and $\delta \phi(j,t_n)$ are random
numbers drawn from independent continuous uniform distributions
defined at every $j$. All distributions are centered at zero and ranging from $-p$ to $p$. 
Note that for $q(j,t_n)$ we take the absolute value of 
the right hand side since we must always have $q(j,t_n)>0$.
We can understand $p$ as the maximal relative variation of $q, \theta$, or $\phi$ 
with respect to their upper bounds. For instance, $p=0.1\%$ means that  
they will change from $t_{n-1}$ to $t_{n}$ by at most $\pm 0.1\%$ of 
their maximal allowed values. For $q$ the maximal value is $1$ while for 
$\theta$ and $\phi$ we have $\pi$. Also, depending on the type of disorder,  
$\delta q(j,t_n), \delta \theta(j,t_n)$, and $\delta \phi(j,t_n)$ are functions only of position, only of time, or of both position and time. In other words, we have, 
respectively, static, dynamic, or fluctuating disorder \cite{vieira2013dynamically,vieira2014entangling,vie18,vie19,vie20}. 

Being more specific, for static disorder we randomly and independently 
change the optimal coin at every 
site $j$ according to the above prescription only once (at $t=0$). For 
dynamic disorder whenever $t=n\tau$, $n=1,2,3,\ldots$, and
$\tau$ a predetermined period, we change every coin 
in the same way, i.e., using the same random number drawn from a given 
uniform distribution. Finally, for fluctuating disorder, whenever $t=n\tau$
we change all coins independently, similar to what we do for static disorder.

\begin{figure}[!htb]
\includegraphics[width=8cm]{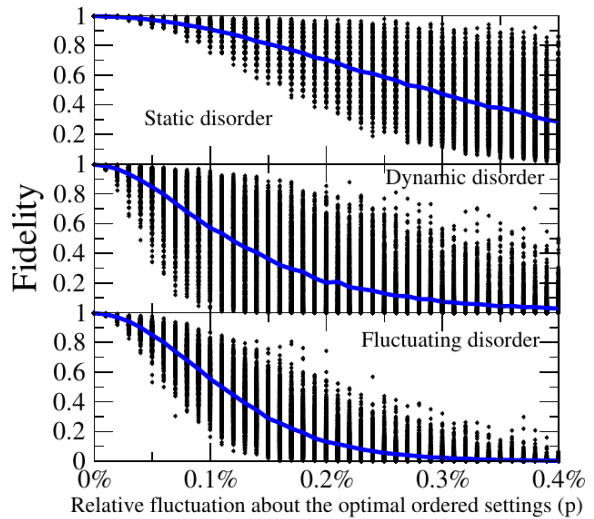}
\caption{(color online). Top to bottom: static, dynamic, and fluctuating disorder
affecting the execution of the multiple station protocol (Fig.~\ref{fig3}). The error
rate $p$ (relative fluctuation about the optimal ordered settings) varies from $0\%$ to $0.4\%$ in increments of $0.01\%$.
For every value of $p$ and type of disorder we implement $1000$ disorder realizations,
represented by the dots in the graphics. The average value for the fidelity is given by the solid-blue lines. The initial spin state for all cases is 
$(\ket{\up}+i\ket{\down})/\sqrt{2}$ and $\tau=10\% t_{M}$, where $t_M$ is the time 
spent by the wave packet to arrive at its last corral with optimal fidelity in the ordered protocol. See text for details.}
\label{fig4}
\end{figure}

In Fig.~\ref{fig4} we show how the multiple station
quantum corralling protocol responds to the above three types of disorder.
We realize that it is least affected by static disorder while fluctuating disorder
is the most severe. Whenever the error rate is below $p=0.05\%$ we always
have an average fidelity of at least $0.8$, even for fluctuating disorder.
This is a quite remarkable result, in particular
if we remember that we are dealing with a thousand-step protocol where 
errors accumulate from one step to another. Also,
the present efficiency is compatible with other state of the art protocols  \cite{crespi2013anderson,gueddana2019can} and we believe we can increase its
efficiency at higher error rates by properly applying quantum error correction strategies at intermediate corrals \cite{aharonov1998fault,aharonov2008fault}.
See also the Appendix \ref{ap1} for a complementary analysis on the effects 
of disorder in the present protocol.

\section{Summary} 

Within the framework of quantum walks 
we proposed a very simple and robust 
way to store and transmit a qubit initially localized in a 
wave package. The present protocol, which we 
dubbed ``quantum corralling'', 
uses only two types of coins, the Hadamard and the $\sigma_x$ coins, 
to effectively generate a dispersionless storage or
transmission of a Gaussian wave package.

The confined or transmitted state, when measured at the right time, showed a high
level of fidelity with the initial one, achieving almost unity fidelity even for walks
of thousands of steps. The protocol worked independently of the initial spin state,
which suggests that it can be used as building blocks 
to the development of dynamical quantum memories if we employ
state of the art implementations of quantum walks \cite{cardano2015quantum,su2019experimental,crespi2013anderson,gueddana2019can}.
Finally, we also envisage the use of the quantum corralling protocol to build 
quantum cargo protocols, where several qubits are sequentially prepared and sent
using single or multiple pathways \cite{sen20}.

\begin{acknowledgments}
GR thanks the Brazilian agency CNPq
(Brazilian National Council for Scientific and Technological Development) for partially funding this research. 
\end{acknowledgments}

\appendix

\section{Analytical proof of Eq.~(\ref{Psit}) of the main text} 
\label{ap0}

To analytically understand the time evolution of a 
Gaussian state when in all sites we have Hadamard coins (Hadamard walk), 
we need to work in the dual $k$-space. 

Defining the two-component vector
\begin{equation}
\Psi(j,t) = 
\begin{bmatrix}
\Psi_{\uparrow}(j,t) \\ \Psi_{\downarrow}(j,t)  
\end{bmatrix} =
\begin{bmatrix}
\left(\langle \uparrow | \langle j | \right) |\Psi(t)\rangle \\ 
\left(\langle \downarrow | \langle j | \right) |\Psi(t)\rangle  
\end{bmatrix},
\label{spinor}
\end{equation}
where $\Psi_{\uparrow(\downarrow)}(j,t)$ is the probability amplitude of finding the qubit 
at position $j$ with spin up (down), the dual $k$-space is defined as follows \cite{nay00},
\begin{equation}
\tilde{\Psi}(k,t) = \sum_j\Psi(j,t)e^{ikj}. 
\label{psik}
\end{equation}
Here the sum runs through all integers from $-\infty$ to $\infty$, 
$k$ is a real number such that $k\in [-\pi,\pi]$,
and  
$\tilde{\Psi}(k,t)$ is a two component vector as well.
The inverse Fourier transform is \cite{nay00}
\begin{equation}
\Psi(j,t) = \frac{1}{2\pi}\int_{-\pi}^{\pi} \tilde{\Psi}(k,t)e^{-ikj}dk.
\label{psij}
\end{equation}

%
%
%

Using this notation, the initial state given in the main text [Eq.~(\ref{state0})] can 
be written as
\begin{equation}
\Psi(j,0) = 
\begin{bmatrix}
\Psi_{\uparrow}(j,0) \\ \Psi_{\downarrow}(j,0)
\end{bmatrix},
\label{psij0}
\end{equation}
where
\begin{eqnarray}
\Psi_{\uparrow}(j,0) &=& f(j)\cos\alpha, \\
\Psi_{\downarrow}(j,0) &=& f(j)e^{i\beta}\sin\alpha, 
\end{eqnarray}
and
\begin{equation}
f(j)=Ae^{-[j^2/(4 s^2)]}/(2\pi s^2)^{1/4}.
\end{equation}
The definition and meaning of $A$ are given in the main text while $s$ 
is the standard deviation of the Gaussian wave packet.

Using Eqs.~(\ref{psik}) and (\ref{psij0}), the initial condition in
the dual $k$-space is
\begin{equation}
\tilde{\Psi}(k,0)=\sum_{j=-\infty}^{+\infty}f(j)e^{ikj} 
\begin{bmatrix}
\cos\alpha \\
e^{i \beta}\sin\alpha
\end{bmatrix}.
\label{Psitil_00}
\end{equation}

For a large enough $s$ we can approximate the above sum for an integral. In this case $A=1$ and
we get
\begin{equation}
\tilde{\Psi}(k,0)\approx (8\pi s^2)^{1/4}e^{-k^2s^2} 
\begin{bmatrix}
\cos\alpha \\
e^{i \beta}\sin\alpha
\end{bmatrix}.
\label{Psitil_0}
\end{equation}

The dynamics in the $k$-space can be deduced by first obtaining the state at
time $t+1$ from the one at $t$ in the position space and then using Eq.~(\ref{psik}). 
Following Ref. \cite{nay00} and adapting the notation to our present problem, 
it is not difficult to see that $\tilde{\Psi}(k,t+1) = M_k\tilde{\Psi}(k,t)$,
where
\begin{equation}
M_k = \frac{1}{\sqrt{2}}
\begin{bmatrix}
e^{ik} & e^{ik} \\ e^{-ik} & - e^{-ik}
\end{bmatrix}.
\end{equation}

Recursively applying this relation we get
\begin{equation}
\tilde{\Psi}(k,t) = (M_k)^t\tilde{\Psi}(k,0).
\label{psirec}
\end{equation}

If we now diagonalize $M_k$ we have
\begin{equation}
M_k=\lambda^{+}_k\ket{u_k^{+}}\bra{u_k^{+}}+\lambda^{-}_k\ket{u_k^{-}}\bra{u_k^{-}},
\label{Mkdiag}
\end{equation}
with eigenvalues 
\begin{equation}
\lambda^{\pm}_k = \pm e^{\pm i\omega_k}, 
\end{equation}
where $\omega_k \in [-\pi/2, \pi/2]$ and 
\begin{equation}
\sin\omega_k  = \frac{\sin k}{\sqrt{2}}.
\end{equation}
The corresponding eigenvectors can be written as
\begin{equation}
\ket{u_k^{\pm}}=\dfrac{
\begin{bmatrix}
1\pm \sqrt{2}e^{i(k\pm \omega_k)} 
\\
1 
\end{bmatrix}}
{\sqrt{2\left(1+\cos^2k\pm \cos k \sqrt{1+\cos^2k}\right)}}.
\label{Phik}
\end{equation}

Therefore, inserting Eq.~(\ref{Mkdiag}) into (\ref{psirec}) we get 
\begin{align}
\tilde{\Psi}(k,t)&=e^{i\omega_k t}\ket{u_k^{+}}\bra{u_k^{+}}\tilde{\Psi}(k,0) \nonumber \\
&+(-1)^te^{-i\omega_kt}\ket{u_k^{-}}\bra{u_k^{-}}\tilde{\Psi}(k,0).
\label{Psikt}
\end{align}

If we now use Eqs.~(\ref{psij}), (\ref{Psitil_0}), (\ref{Phik}), and (\ref{Psikt})
we obtain
\begin{align}
\Psi(j,t)&\approx \int_{-\pi}^{+\pi}\frac{dk}{2\pi} e^{-ikj} \left[(8\pi s^2)^{1/4}e^{-k^2s^2}  \right]
\times\nonumber\\
&\left\{
e^{i\omega_kt}g_+(k) 
\begin{bmatrix}
1+ \sqrt{2}e^{i(k+\omega_k)}\\
1
\end{bmatrix}\right.\nonumber\\
&\left.+(-1)^te^{-i\omega_kt}g_-(k)
\begin{bmatrix}
1- \sqrt{2}e^{i(k-\omega_k)}  \\
1
\end{bmatrix}
\right\},
\label{Psixt1}
\end{align}
where
\begin{equation}
g_{\pm}(k)=\frac{e^{i \beta}\sin \alpha + \left(1 \pm \sqrt{2}e^{-i(k \pm \omega_k)} \right) \cos \alpha}
{2\left(1+\cos^2k \pm \cos k \sqrt{1+\cos^2k}\right)}.
\label{fk}
\end{equation}

We now employ once more the assumption that $s$ is sufficiently large and also assume that $t$ is not too
big. Since a large $s$
means a very narrow wave packet in the $k$-space centered about $k=0$, we can 
extend the above integration from $-\infty$ to $\infty$ and make the following approximations:
\begin{eqnarray}
\omega_k t &\approx& \frac{k}{\sqrt{2}}t + \mathcal{O}(k^3)t, \label{aprox1} \\
e^{-i(k \pm \omega_k)} &\approx& 1 + \mathcal{O}(k), \label{aprox2}\\
g_{\pm}(k) &\approx& \frac{e^{i\beta}\sin\alpha + (1\pm\sqrt{2})\cos\alpha}{2(2\pm\sqrt{2})} + \mathcal{O}(k). 
\label{aprox3}
\end{eqnarray}

Inserting Eqs.~(\ref{aprox1})-(\ref{aprox3}) into (\ref{Psixt1}) and carrying out the integration we get
%
%
\begin{eqnarray}
\Psi(j,t)&\approx&
h_+(\alpha,\beta) \ket{R}f(j-t/\sqrt{2}) \nonumber \\
&+&(-1)^{t} h_-(\alpha,\beta)
\ket{L}f(j+t/\sqrt{2}), 
\label{Psitil_8}
\end{eqnarray}
where
\begin{eqnarray}
h_{\pm}(\alpha,\beta) = \frac{e^{i \beta}\!\sin \alpha\! +\!\left(1\pm \sqrt{2} \right)\!\cos \alpha}
{\sqrt{2\left(2\pm \sqrt{2}\right)}}
\end{eqnarray}
and
\begin{eqnarray}
|R\rangle  &=&  \frac{1}{\sqrt{2\left(2+ \sqrt{2}\right)}}
\begin{bmatrix}
 1 \!+\!\sqrt{2}\\
1
\end{bmatrix}, \\
|L\rangle  &=&  \frac{1}{\sqrt{2\left(2- \sqrt{2}\right)}}
\begin{bmatrix}
 1 \!-\! \sqrt{2} \\
1 
\end{bmatrix}.
\end{eqnarray}
Here $\ket{R}$ and $\ket{L}$ are normalized orthogonal states.

If we now define the two orthogonal states
\begin{eqnarray}
|\psi_{R}\rangle &=& h_+(\alpha,\beta)
|R\rangle, \\
|\psi_{L}\rangle &=& h_-(\alpha,\beta)
|L\rangle,
\end{eqnarray}
we immediately see that we can write Eq.~(\ref{Psitil_8}) as
\begin{eqnarray}
\Psi(j,t) &\approx& \braket{R|\psi_R}|R\rangle
f(j-t/\sqrt{2}) \nonumber \\
&&+(-1)^{t} \braket{L|\psi_L}|L\rangle
f(j+t/\sqrt{2}). 
\label{Psitil_9}
\end{eqnarray}

Now, working in the $\{|R\rangle,|L\rangle\}$ basis and making the following identification,
\begin{equation}
\Psi(j,t) = 
\begin{bmatrix}
\left(\langle R | \langle j | \right) |\Psi(t)\rangle \\ 
\left(\langle L | \langle j | \right) |\Psi(t)\rangle  
\end{bmatrix} 
,
\label{spinor2}
\end{equation}
we have that
\begin{eqnarray} 
\ket{\Psi(t)}&=&\ket{\psi_R}\otimes \sum_jf(j-t/\sqrt{2})|j\rangle \nonumber \\
&& + (-1)^t\ket{\psi_L}\otimes\sum_jf(j+t/\sqrt{2})|j\rangle
\label{Psitmain}
\end{eqnarray}
is the same as Eq.~(\ref{Psitil_9}). And the proof is finished by noting that Eq.~(\ref{Psitmain}) is Eq.~(\ref{Psit}) of the main text.

\section{Influence of the wave package's
initial dispersion in position on the protocol's efficiency} 
\label{apB}

\begin{figure}[!htb]
\includegraphics[width=8cm]{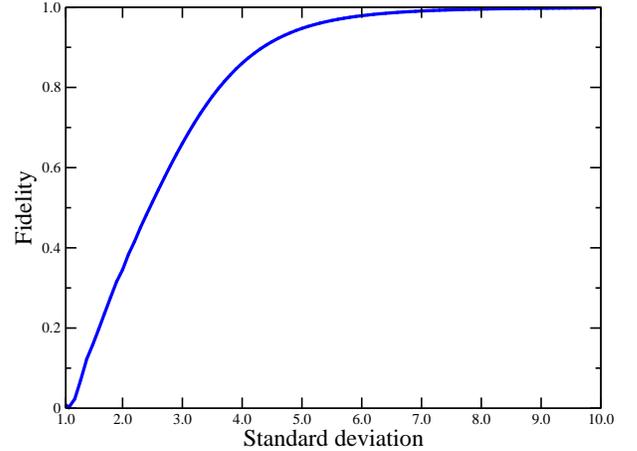}
\caption{(color online). The data above were obtained using the same setting of Fig.~\ref{fig2} of the main text (single shot herding). The only difference is that we 
now fix our attention on one initial internal state, $(|\up\rangle - i|\down\rangle)/\sqrt{2}$, and change the standard deviation $s$ of the wave package from $s=10$, the value used in Fig.~\ref{fig2}, down to $s=1.0$. For every value of $s$, we evolve the system 
according to the single shot herding protocol and we measure the system
at the optimal time given in Fig.~\ref{fig2}. This is the state used to compute the 
fidelity with the initial state for a given value of $s$.}
\label{fig6}
\end{figure}

Here we give a more quantitative view of the efficiency of the protocol when the standard deviation in position of the initial Gaussian wave package decreases. As can be seen analyzing 
Fig.~\ref{fig6}, the narrower the Gaussian wave package (lower the standard deviation), the 
less efficient is the protocol. The plot in Fig.~\ref{fig6} is made for a particular 
initial spin state but the general trend is similar for other initial spin states. 
For a standard deviation $s\geq 5.0$, we always get a fidelity of transmission at least of the order of $0.9$. As we further  decrease $s$, the fidelity rapidly decreases and the
present protocol is no longer the best option to transmit localized states.     

\section{More on disorder} 
\label{ap1}

Our goal here is to investigate the response of the multiple station protocol to the following two scenarios of disorder.
First, we want to know the efficiency of the protocol when the bias $q(j,t)$ of the coins are subjected to disorder 
while the phases are not. Second, what happens if now the phases are affected by disorder and the bias of the coins is
unaffected.

\begin{figure}[!htb]
\includegraphics[width=8cm]{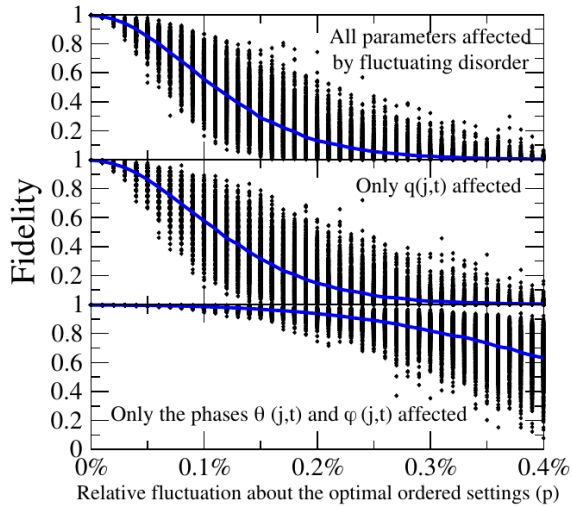}
\caption{(color online). 
The upper panel is exactly the lower panel of Fig.~4 of the main text, where fluctuating disorder is introduced in the 
multiple station corralling protocol, affecting all parameters of the coins, namely, $q(j,t)$, $\theta(j,t)$, and
$\phi(j,t)$. We also employ the same notation in the middle and lower panels.
The data of the other panels were computed using the same settings and number of disorder realizations for each $p$ as given in Fig.~4 
of the main text, with the following modifications.  
Middle panel: Fluctuating disorder acting only on the parameters $q(j,t)$ defining 
the Hadamard and $\sigma_x$ coins.  
The phases are not affected by disorder. Lower panel: Fluctuating disorder acting only on the phases, the bias
$q(j,t)$ of the coins are not affected.}
\label{fig5}
\end{figure}

Looking at Fig.~\ref{fig5} we realize that the system is barely affected when disorder is present only in the phases. The 
relevant parameter which determines the whole fate of the 
protocol in the presence of disorder is the bias $q(j,t)$. Comparing 
the upper panel with the middle one, we see that they lead to almost the same fidelities. The lower panel shows that 
when only the phases are subjected to disorder, we can have a much greater value of error $p$ without appreciably affecting 
the efficiency of the protocol.

We also checked a possible decrease in the efficiency of the protocol 
when we measure the transmitted quantum state at a different time than the 
optimal one predicted by the clean model. The first thing worth noticing is the fact that if we measure the state one time step before or 
after the right time $t_M$ we get zero fidelity. This is related to the $(-1)^t$ relative phase between the two split wave packages,
as depicted in Eq.~(\ref{Psit}) of the main text [Eq.~(\ref{Psitmain}) here]. Actually, if we measure the wave package at odd times
we will always get a relative $\pi$ phase, i.e., $(-1)^{t_{odd}}=-1$. In this case we have to apply an appropriate phase flip gate
to compensate for this phase. For even times, but not too distant from the correct 
measuring time, we get very high fidelities, almost as high as if we had measured at the right time. With that in mind, we tested what would happen if we deviate about $t_M$, detecting the state before or after
the right time. We observed that for deviations of the order $\pm 10\%$ about $t_M$, 
no appreciable reduction in the fidelity occurred. We still
get in this scenario an average fidelity greater than $0.9$.

Finally, we also investigated how fluctuations about the right time to change the Hadamard coin to a $\sigma_x$ coin (closing the gate) or vice-versa affected the protocol. The decrease in the efficiency of the protocol was negligible to deviations of the order 
$\pm 10\%$ about the correct
time to switch one coin to the other. This comes about because the switching of the coins occurs several standard deviations away from the 
center of the wave package.

\section{Description of the accompanying videos} 
\label{ap2}

The file ``single\_shot\_corralling.mp4'' is the animation of the single shot protocol as given in Fig.~2 of the 
main text using the state $(|\up\rangle+i|\down\rangle)/\sqrt{2}$ as the initial spin state. 
For every integer $t$, from $t=0$ to  $t=995$ steps, we have computed the probability distribution
and then animated those $996$ frames. The green vertical bars mark the lattice sites
where a Hadamard coin was changed to a 
$\sigma_x$ coin (closing the gate).

The file ``multiple\_station\_corralling.mp4'' is the animation of the multiple station protocol as given in Fig.~3 of the 
main text using the state $(|\up\rangle+i|\down\rangle)/\sqrt{2}$ as the initial qubit state. 
For every integer $t$, from $t=0$ to  $t=1566$, we have computed the probability distribution
and then animated those $1567$ frames. The green vertical bars mark where a Hadamard coin was changed to a 
$\sigma_x$ coin (closing the gate).

\end{document}